\documentclass[final]{svjour2}
\usepackage{graphicx}
\usepackage{rotating}
\usepackage{amssymb}
\usepackage{mathptmx}
\usepackage[numbers]{natbib}
\makeatletter
\journalname{Journal of Low Temperature Physics}

\bibpunct{}{}{,}{s}{}{,}

\begin{document}

\newcommand{\hdblarrow}{H\makebox[0.9ex][l]{$\downdownarrows$}-}
\title{On Larkin-Imry-Ma State of $^3$He-A in Aerogel}

\author{G.E. Volovik}

\institute{Low Temperature Laboratory, 
Helsinki University of Technology,
P.O.Box 2200, FIN-02015 HUT, Finland, and
L.D. Landau Institute for Theoretical Physics, 119334
 Moscow, Russia\\ Tel.: 358-9-4512963\\ Fax: 358-9-4512969\\
\email{volovik@boojum.hut.fi}}

\date{01.10.2007}

\maketitle


\begin{abstract}

Superfluid $^3$He-A shares the properties of spin nematic and chiral orbital ferromagnet. Its order parameter is characterized by two vectors  $\hat\textbf{d}$ and $\hat\textbf{l}$. This doubly  anisotropic superfluid, when it is confined in aerogel,   represents the most interesting example of a system  with continuous symmetry in the presence
of random anisotropy disorder. We discuss the Larkin-Imry-Ma state, which is characterized
by the short-range orientational order of the vector $\hat\textbf{l}$, while the long-range orientational order is destroyed by the collective action of the randomly oriented aerogel strings. On the other hand, sufficiently large regular anisotropy produced either by the deformation of the aerogel or by applied superflow   suppresses  the Larkin-Imry-Ma effect leading to the uniform orientation of $\hat\textbf{l}$. The interplay of regular and random anisotropy allows us to study many different effects.


\end{abstract}

\section{Introduction}

\begin{figure}
\centerline{\includegraphics[width=1.0\linewidth]{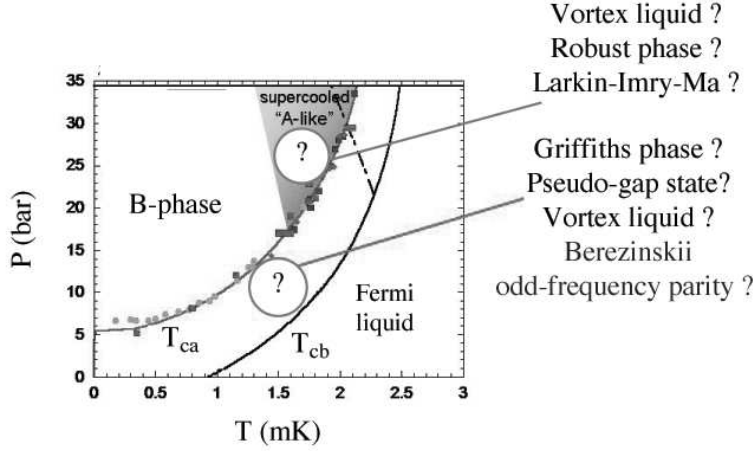}}
\caption{Two main problems for superfluid $^3$He in aerogel: (1) What is the state of $^3$He between  two lines of  superfluid phase transition? What is the state of $^3$He in the so-called A-like phase?}  
\label{TwoProblems} 
\end{figure}

NMR experiments on liquid  $^3$He  confined in aerogel demonstrate  a rich but still unexplained life in two regions of the phase diagram in Fig. \ref{TwoProblems}: (i) in the region between the lines $T_{\rm cb}$ and  $T_{\rm ca}$ of superfluid phase transition in bulk liquid and   in aerogel correspondingly \cite{Between} (in principle, Berezinskii odd-frequency pairing state is possible there \cite{Tanaka}); and (ii)  in the region of the so-called  A-like phase which is metastable in the absence of external magnetic field \cite{Dmitriev,BunkovDeformed,KunimatsuQFS}. Here we concentrate on the A-like phase, which will be discussed in terms of the Larkin-Imry-Ma effect occurring in the anisotropic  $^3$He-A  superfluid when it is confined in aerogel.

\subsection{Doubly anisotropic superfluid  $^3$He-A}

Superfluid $^3$He-A shares the properties of spin nematic and chiral orbital ferromagnet. Its order parameter is characterized by two vectors  $\hat\textbf{d}$ and $\hat\textbf{l}$ \cite{VollhardtWolfle}:
\begin{equation}
 A_{\mu i}=\Delta e^{i\phi}\hat d_\mu(\hat m_i+i\hat n_i)~~,~~ \hat\textbf{l} =\hat\textbf{m}\times \hat\textbf{n}~.
\label{OrderParameter}
\end{equation}
Unit vector $\hat\textbf{d}$ marks direction of anisotropy axis in spin space, and is responsible for anisotropy of spin susceptibility; unit vector $\hat\textbf{l}$ marks the direction of the  orbital  angular momentum of Cooper pairs and simultaneously the direction of axis of the orbital anisotropy and it is responsible for anisotropy of superfluid density:
\begin{equation}
F_s+F_o+F_{so}= \frac{(\textbf{S}\cdot \hat\textbf{d})^2 }{2\chi_{\parallel}}+\frac{(\textbf{S}\times\hat\textbf{d})^2}{2\chi_{\perp}}+\frac{\rho_{s\parallel}}{2}(\textbf{v}_s\cdot \hat\textbf{l})^2 +\frac{\rho_{s\perp}}{2}(\textbf{v}_s\times\hat\textbf{l})^2-g_D(\hat\textbf{d}\cdot \hat\textbf{l})^2.
\label{Anisotropy}
\end{equation}
Here $\textbf{S}$ is spin density in the applied magnetic field, $S_\alpha=\chi_{\alpha\beta}H_\beta$; $ \textbf{v}_s$ is the velocity of superfluid mass current $j_s^i=\rho_s^{ij}v_s^j$. Because of the chiral nature of the orbital order parameter $\hat\textbf{m}+i\hat\textbf{n}$, the superfluid velocity
and its vorticity 
\begin{equation}
 \textbf{v}_s=\frac{\hbar}{2m}\left(\nabla\phi + \hat m_i\nabla\hat n_i\right)  ~~,~~
 \nabla\times \textbf{v}_s=\frac{\hbar}{4m} e^{ijk} \hat l_i \nabla\hat l_j\times\nabla\hat l_k~,
\label{VelocityVorticity}
\end{equation}
can be generated by the texture of the $\hat\textbf{l}$-vector \cite{MerminHo}.
The last term in Eq.(\ref{Anisotropy}) describes a tiny spin-orbit coupling $F_{so}$ between $\hat\textbf{d}$ and $\hat\textbf{l}$, which is however very important for the NMR measurements since it produces the frequency shift. 

\begin{figure}
\centerline{\includegraphics[width=1.0\linewidth]{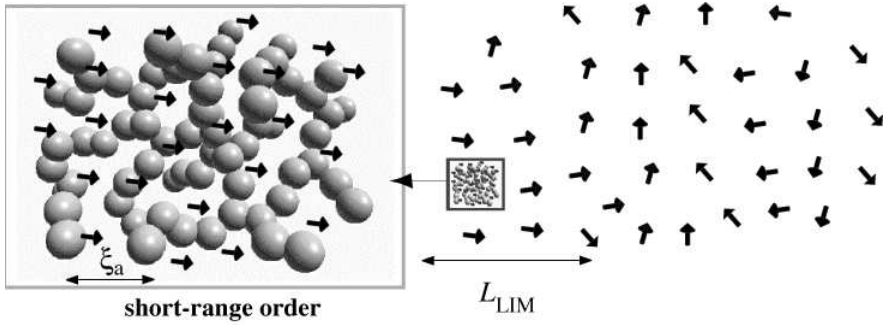}}
\caption{While the short-range order is well defined, the collective orientational effect of aerogel strings destroys the long-range order on a macroscopic Larkin-Imry-Ma scale $L_{\rm LIM}$}  
\label{fig:texture} 
\end{figure}

This doubly  anisotropic superfluid, when it is confined in aerogel,   represents the most interesting example of a system  with continuous symmetry in the presence
of random anisotropy. 
For such systems the surprizing conjecture was made by Imry and Ma  \cite{ImryMa} that even a weak disorder may destroy   the long-range  orientational order (LROO). This is the counterpart of the effect of collective pinning in superconductors  predicted by Larkin  \cite{Larkin}, in which  weak impurities destroy the long-range  translational  order of the Abrikosov vortex  lattice.
Recent NMR experiments allow us to discuss
the state of  $^3$He-A in aerogel in terms of the Larkin-Imry-Ma (LIM) state. The LIM state in $^3$He-A is characterized by short-range orientational order of the orbital vector $\hat\textbf{l}$, while the LROO  is destroyed by the collective action of the randomly oriented aerogel strings (Fig. \ref{fig:texture}).  The order of magnitude theoretical estimations based on the model of aerogel in Sec. \ref{Model} and the NMR data suggest that in the aerogel samples under investigation,  the LIM length $L_{\rm LIM}$ at which the LROO is destroyed is of order of a micron. 

Though $L_{\rm LIM}$ is much bigger than the microscopic scales of aerogel strands, the orientational disorder may have pronounced effects.  In particular, it may be the reason of existence of a small region of equilibrium A-phase in zero magnetic field \cite{Vicente}. For NMR experiments it is important that
$L_{\rm LIM}$ is smaller than  dipole length $\xi_D\sim 10~\mu$m  characterizing  spin-orbit coupling between $\hat\textbf{l}$ and $\hat\textbf{d}$ in Eq.(\ref{Anisotropy}). This leads to anomalously  small values of the observed transverse NMR frequency shift and longitudinal NMR frequency \cite{Dmitriev} as we shall discuss in Sec. \ref{Experiment}.  

A regular anisotropy may suppress the LIM effect, and this does take place when one applies a uniaxial deformation to the aerogel sample (Sec. \ref{SectionRegular}). A linear deformation of about 1$\%$ leads to the restoration of the uniform orientation of the  $\hat\textbf{l}$-vector in the sample \cite{BunkovDeformed}. As a result the value 
of the transverse NMR frequency shift is  enhanced by an order of magnitude compared to that in the LIM state. If the deformation leads to orientation of  $\hat\textbf{l}$  along the magnetic field $\textbf{H}$, the transverse NMR frequency shift becomes negative. These are the observations which support the interpretation of the A-phase like state in aerogel as the disordered  LIM state.

Open problems related to the disordered LIM state including the superfluid properties of the LIM state
are  discussed in Sec. \ref{Superfluidity}. Superfluidity in LIM state could be rather unusual because, according to the Mermi-Ho relation in Eq.(\ref{VelocityVorticity}) between  $\hat\textbf{l}$ and $\textbf{v}_s$, the LIM texture generates superfluid vorticity. As a result, the disordered LIM state can be represented as  a system of randomly distributed skyrmions -- vortices with continuous cores. The characteristic distance between the vortex-skyrmions and the size of their cores are determined by the LIM length  $L_{\rm LIM}$. In principle, the superfluid density of  $^3$He-A in aerogel may depend on the ability of the aerogel to pin the randomly distributed vortex-skyrmions.

\section{Theory of  Larkin-Imry-Ma effect}
\label{SectionTheory}

\subsection{Larkin-Imry-Ma state in the model of random cylinders}
\label{Model}

\begin{figure}
\centerline{\includegraphics[width=1.0\linewidth]{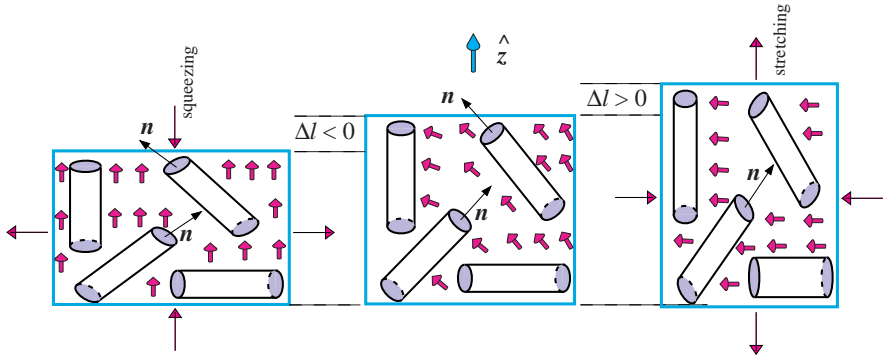}}
\caption{Uniaxial deformation of cylindrical sample of aerogel. Squeezing  (stretching) of cylinder along its axis $z$ leads to global anisotropy for $\hat\textbf{l}$  with   easy axis  along $z$ (easy plane normal to $z$)}  
\label{fig:RegularAnisotropy} 
\end{figure}

We use a simple model of aerogel as a  system of randomly oriented cylindrical strands (see Refs.\cite{Volovik1,SurFom} and Fig. \ref{fig:RegularAnisotropy} {\it middle}) with diameter $\delta\sim 3$ nm and length $\xi_{\rm a}\sim $ 20 nm, which is also the distance between  strands. Since $\delta$ is much smaller than the superfluid coherence length $\xi_0$,  the theory of Rainer and Vuorio  for the microscopic body immersed in $^3$He-A \cite{RainerVuorio} can be applied. According to \cite{RainerVuorio}
the  $\hat\textbf{l}$-vector remains uniform outside the body (see Fig. \ref{fig:texture} {\it left}), and the orientational energy acting from $\hat\textbf{l}$ on a cylinder $i$ with the direction of axis  
along $\hat\textbf{n}_i$ is 
\begin{equation}
E_i =E_{\rm a}(\hat\textbf{l}\cdot\hat\textbf{n}_i)^2 ~, ~
E_{\rm a}\sim  \frac{\Delta^2}{T_c}  k_F^2 \xi_{\rm a}\delta>0~.
\label{OrientationalEnergy}
\end{equation}
Here $k_F$ is Fermi momentum;  the parameter $E_{\rm a}$ is positive as follows from \cite{RainerVuorio}.

For the infinite system of cylinders,  the average orienational effect of the random cylinders on the $\hat\textbf{l}$-vector is absent, if the $\hat\textbf{l}$-vector is kept uniform.  However, there is  a collective effect of many cylinders, which makes the $\hat\textbf{l}$-vector inhomogeneous on large scales. Let us consider  the box of size $L\times L\times L$,  which contains a large but finite number $N\sim L^3/\xi_{\rm a}^3\gg 1$ of randomly oriented cylinders and still has a uniform orientation of the $\hat\textbf{l}$-vector. Due to fluctuations of  orientational energies of randomly oriented cylinders, the $\hat\textbf{l}$-vector will find the preferred orientation in the box, with the energy gain proportional to $N^{1/2}$.
The corresponding negative  energy density is determined by the variance of the energy of random anisotropy:
\begin{equation}
E_{ran}\sim  - \left< \sum_{i=1}^{N}\left(E_i -\left<  E\right>\right)^2\right> ^{1/2}L^{-3} \sim -N^{1/2}  E_{\rm a}L^{-3} \sim  - E_{\rm a} \xi_{\rm a}^{-3/2}L^{-3/2}  ~.
\label{RandomAnisotropyEnergy}
\end{equation}
The neighboring boxes prefer  different orientations of $\hat\textbf{l}$, as a result the effect of  orientation by the aerogel strands will be opposed  by the gradient energy $E_{grad}  \sim K(\nabla\hat\textbf{l})^2\sim K/L^2$. The optimal size of a box with a uniform orientation of $\hat\textbf{l}$ within the box -- the LIM length $L_{\rm LIM}$ at which LROO  is destroyed -- is found from  competition  between the energy  of random anisotropy and the gradient energy:
\begin{equation}
E_{ran}+E_{grad}  \sim  - E_{\rm a} \xi_{\rm a}^{-3/2}L^{-3/2} + KL^{-2}
 ~.
\label{Random+GradEnergy}
\end{equation}
Minimization of Eq.(\ref{Random+GradEnergy}) gives
\begin{equation}
L_{\rm LIM}\sim \frac{K^2\xi_{\rm a}^3}{E_{\rm a}^2} \sim \xi_{\rm a}\frac{\xi_0^2}{\delta^2}~.
\label{LIMlength}
\end{equation}
Here  we used Eq.(\ref{OrientationalEnergy}) for $E_{\rm a}$; and  estimated the rigidity of  
$\hat\textbf{l}$  as $K\sim (k_F^3/m)(\Delta^2/T_c^2)$. 

Eq.(\ref{LIMlength}) is valid for thin strings, 
$\delta \ll \xi_0$. This corresponds to collective pinning:  $\hat\textbf{l}$ is not pinned by individual strings, has a short-range orientational order and is perfectly uniform at the  aerogel scale $\xi_{\rm a}$. The  LROO is lost at  scale $L_{\rm LIM}\gg  \xi_{\rm a}$ due to collective effect of many strings (Fig. \ref{fig:texture} {\it right}). With $\xi_{\rm a}\sim \xi_0\sim 20$ 
nm and $\delta \sim 3$ nm, the length
$L_{\rm LIM}\sim 1~\mu$m.    Strong pinning would occur in case of thick strings, $\delta > \xi_0$, when $\hat\textbf{l}$ is oriented by each string due to  boundary conditions for $\hat\textbf{l}$ at the surface of a string, and the LROO  is destroyed at the scale of $\xi_{\rm a}$.

\subsection{LIM effect killed by regular anisotropy}
\label{SectionRegular}

Squeezing of a cylindrical sample of aerogel along its axis ($u_{zz}\sim \Delta l/l<0$, Fig. \ref{fig:RegularAnisotropy} {\it left})  makes $z$ the easy axis  for $\hat\textbf{l}$, while  stretching ($u_{zz}\sim\Delta l/l>0$, Fig. \ref{fig:RegularAnisotropy} {\it right}) produces the easy plane anisotropy. 
In general,  the deformation $u_{ik}$ of aerogel  induces  regular anisotropy for the $\hat\textbf{l}$-vector with the density of orientational energy:
\begin{equation}
E_{reg}= \left< \sum_{i}\left(E_i -\left<  E\right>\right)\right> =\kappa \left(u_{ik} -\frac{1}{3}u_{ll}\delta_{ik}\right)\hat l_i\hat l_k ~~,~~\kappa \sim E_{\rm a}  \xi_{\rm a}^{-3}~.
\label{RegularAnisotropyEnergy}
\end{equation}
Comparing this energy  with the LIM energy (in Eq.(\ref{Random+GradEnergy}) at $L=L_{\rm LIM}$), one finds that the ordered state with $\hat\textbf{l}$  oriented   along the easy axis $\hat\textbf{z}$ becomes more energetically favorable than the LIM state when the squeezing is sufficiently large:
\begin{equation}
\frac{\vert \Delta l\vert}{l}>  \left(\frac{\xi_{\rm a}}{L_{\rm LIM}}\right)^{3/2} ~.
\label{CriticalSqueezing}
\end{equation}
In the weak pinning limit of $^3$He-A in aerogel the LIM effect is so subtle, that  the critical value of the deformation at which the first order phase transition from the LIM state to the uniform state occurs is rather small; $\Delta l/l \sim 10^{-3}-10^{-2}$.  For large stretching the polar phase is predicted to appear in  vicinity of $T_{\rm ca}$ \cite{Aoyama2}.
Experimentally, squeezing by about 1$\%$ is enough to get the uniform $\hat\textbf{l}$ texture \cite{BunkovDeformed}. 
On the other hand, recent experiments \cite{DmitrievKrasnikhin} in which compression has been applied to compensate intrinsic  deformations, indicate that when the deformation becomes below 1$\%$, the initial uniform $\hat\textbf{l}$ texture is transformed to the disordered state.

\section{Experiment}
\label{Experiment}

\subsection{Estimation of Larkin-Imry-Ma length from NMR}

\begin{figure}
\centerline{\includegraphics[width=1.0\linewidth]{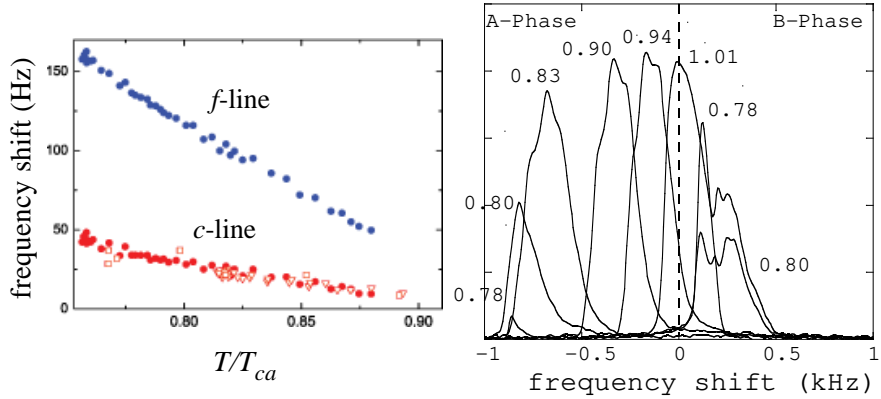}}
\caption{{\it left}: Temperature dependence of positive frequency shift for two lines observed in NMR spectrum on $^3$He-A in aerogel sample according to Ref.  \protect\cite{Dmitriev} ($P=29.3$ bar, $H=224$ G). {\it right}:  Negative frequency shift  in the NMR spectrum observed in $^3$He-A in deformed aerogel according to Ref.  \protect\cite{BunkovDeformed}   ($P=29.3$ bar, $H=290$ G) }  
\label{fig:TwoLines} 
\end{figure}

Fig. \ref{fig:TwoLines}  demonstrates results of the NMR experiments on $^3$He-A in conventional aerogel sample \cite{Dmitriev} ({\it left}), and in a squeezed aerogel \cite{BunkovDeformed} ({\it right}). In the latter, the transverse NMR line has  negative frequency shift corresponding to the uniform $\hat\textbf{l}$-vector  oriented along the magnetic field $\textbf{H}\parallel\hat\textbf{z}$.  

In Ref.  \cite{Dmitriev} two NMR lines have been observed in the transverse NMR spectrum in Fig. \ref{fig:TwoLines}  ({\it left}). The line with a larger frequency shift,  denoted as $f$-line,  can be removed after application of the 180$^\circ$ pulse while cooling through $T_{ca}$ (the superfluid transition temperature in aerogel). The salient feature of the $c$-line which survives after the 180$^\circ$ pulse is the anomalously small frequency shift. It is about 30 times smaller than the magnitude of the negative frequency shift observed in Refs.\cite{BunkovDeformed,KunimatsuQFS} in the aerogel sample with the uniform $\hat\textbf{l}$-vector (we take into account that the frequency shift is $\propto 1/H$ with   $H=224$ G in Ref.\cite{Dmitriev} and $H=290$ G  in Ref.  \cite{BunkovDeformed}). This large difference can be easily understood if  one identifies the $c$-state (the state with a single $c$-line in the spectrum) as the LIM state. 

Properties of NMR on the LIM state depend on ratio between $L_{\rm LIM}$ and   the dipole length $\xi_D\sim 10~\mu$m of  spin-orbit coupling between $\hat\textbf{l}$ and $\hat\textbf{d}$. The orbital vector $\hat\textbf{l}$ will be  locked with the spin-space vector $\hat\textbf{d}$ if $L_{\rm LIM}>\xi_D$  and unlocked from $\hat\textbf{d}$ if $L_{\rm LIM}<\xi_D$. 
The NMR results are in favor of the almost completely unlocked case, since in the limit $L_{\rm LIM}/ \xi_D \ll 1$, the $\hat\textbf{l}$-texture  is fully disordered, $\left< l_z^2\right>=1/3$, and  the frequency shift is zero \cite{Volovik2}.  The nonzero but  relatively small magnitude of the frequency shift compared to the negative frequency shift,  $\Delta \omega_{c-{\rm line}}\sim \vert \Delta \omega_{{\rm deformed}} \vert/30$, can be ascribed to  the $(L_{\rm LIM}/\xi_D)^2$ correction. This allows us to estimate 
$L_{\rm LIM}\sim 1~{\mu}m$, which is in a reasonable agreement with Eq.(\ref{LIMlength}).

\subsection{Tipping angle dependence of NMR}

\begin{figure}
\centerline{\includegraphics[width=1.0\linewidth]{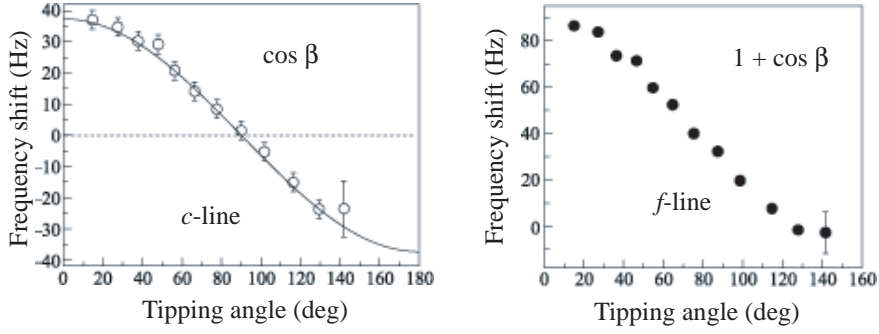}}
\caption{Tipping angle dependence of the frequency shift of transverse NMR in the $c$ and $f$ states of $^3$He-A in aerogel according to Ref.  \protect\cite{Dmitriev} Experiments with compensated deformation demonstrated the $(1+\cos\beta)$ behavior similar to the $f$-line  \protect\cite{DmitrievKrasnikhin}}  
\label{fig:Tipping} 
\end{figure}

Lines $c$ and $f$  in the spectrum of transverse NMR in   $^3$He-A in aerogel demonstrate different dependence of  frequency shift  on the tipping angle $\beta$ in Fig. \ref{fig:Tipping}.  This can be compared with the theoretical frequency shift in the limit of the almost completely disordered state, when  $L_{\rm LIM}/ \xi_D \ll 1$:
\begin{equation}
\frac{ \Delta\omega(\beta)}{ \Delta\omega_{\rm max}} =b\cos\beta+a(1+\cos\beta) ~.
\label{TippingDependence}
\end{equation}
Here $\Delta\omega_{\rm max}$ is the maximum possible frequency shift occurring in the uniform $\hat\textbf{l}$-field, it   coincides with the magnitude of the negative frequency shift when $\hat\textbf{l}\parallel \textbf{H}$;  the parameter $a\ll 1$ and $b\ll 1$ are nonzero due to the deviations from the full disorder caused by spin-orbit interaction \cite{Volovik2}:
\begin{equation}
b=\frac{1}{2} \left(1-3\left<l_z^2\right>\right)~~,~~a=\frac{1}{6} \left(1-2\left<\sin^2\Phi\right>\right)~,
\label{DeviationFromDisorder}
\end{equation}
and $\Phi$ is the angle between the  components  $\hat\textbf{l}_\perp$ and $\hat\textbf{d}_\perp$, which are transverse to magnetic field $\textbf{H}\parallel \hat\textbf{z}$. Parameters $a$ and $b$ are zero in the limit when disorder is strong compared to the spin-orbit interaction, i.e. when $L_{\rm LIM}/ \xi_D\rightarrow 0$, and are proportional to $(L_{\rm LIM}/ \xi_D)^2$ with pre-factors depending on the type of disorder. 

The disorder   can be described in the following  phenomenological way.
 In the $c$-state the transverse component $\hat\textbf{l}_\perp$ of the vector $\hat\textbf{l}$ is random, while the spin-space vector $\hat\textbf{d}$ is perpendicular  to $\textbf{H}$ and is regular. This gives $a=0$. On the other hand the magnetic field $\textbf{H}\parallel \hat\textbf{z}$, via the  vector $\hat\textbf{d}$, produces a weak easy-plane  anisotropy for the  $\hat\textbf{l}$-vector. As a result, one has $b>0$.
The $f$-line represents regions with the $f$-state where  the  $\hat\textbf{l}$-vector is fully random, and thus $b=0$, while  $\hat\textbf{d}$ and $\hat\textbf{l}_\perp$ are not fully independent, producing $a>0$. 
  
  However, it is not clear what is the microscopic background for such behavior of the parameters, and the detailed numerical simulations of the structure of different possible disordered states of $^3$He-A in aerogel are required. Another interpretation of the $\beta$-dependence of the two NMR lines in terms of the modified robust phase has been proposed by Fomin \cite{Fomin}.

\section{Discussion: superfluid properties of LIM state}
\label{Superfluidity}

In chiral anisotropic superfluids the LIM effect should influence the superfluid properties of the system.
According to the Mermin-Ho relation in Eq.(\ref{VelocityVorticity}),  circulation of superfluid velocity along the closed path $L$ is expressed in terms of the surface integral over the $\hat\textbf{l}$-field
\begin{equation}
N(L)=\frac{1}{\kappa}\oint_L d\textbf{r}\cdot \textbf{v}_s=\frac{1}{4\pi}e^{ijk}\int_{S} dS_k ~\hat\textbf{
l}\cdot \left({\partial \hat\textbf{l}\over\partial x^{i}} \times {\partial
\hat\textbf{l}\over\partial x^{j}}\right)~.
\label{Circulation}
\end{equation}
Here we normalized the circulation to the circulation quantum in $^3$He, $\kappa=\pi\hbar/m$; the circulation number  $N$ is not necessarily integer  in chiral  superfluids. 

If we ignore for a moment that the $\hat\textbf{l}$-field is the part of the order parameter triad  $\hat\textbf{m}$,  $\hat\textbf{n}$ and  $\hat\textbf{l}$ in Eq.(\ref{OrderParameter}) and consider $\hat\textbf{l}$ as completely independent field, then the variance of $N$  for sufficiently large contours is 
 \begin{equation}
 \left<N^2(L)\right>\sim \frac{L^2}{L^2_{\rm LIM}}~~,~~L\gg L_{\rm LIM}~.
\label{VarianceN}
\end{equation}
This corresponds to randomly distributed skyrmions -- vortices with continuous cores of the size $L_{\rm LIM}$, which is also the mean distance between the skyrmions.  Such a state resembles the plasma of vortices above the Berezinskii-Kosterlitz-Thouless transition, where vortices destroy superfluidity. This suggests that in aerogel,  while the local superfluid density $\rho_s$ on the scales below $L_{\rm LIM}$ is of the same order as in $^3$He-B, it could be reduced and even become zero at large scales.

 Experiments on $^3$He-A in aerogel  \cite{Nazaretski,Fisher} demonstrate that  $\rho_s\neq 0$  though it is somewhat smaller  as compared to $\rho_s$ in $^3$He-B.
  This shows that  in the consideration of the LIM effect, we cannot ignore the energy of superflow --  the energy  related to the gradient  $\hat\textbf{m}\cdot\nabla \hat\textbf{n}$, which is the superfluid velocity according to  Eq.(\ref{VelocityVorticity}).  It was stated that the energy of superflow is not relevant for the LIM effect \cite{Aoyama1}. However, it should suppress the abundance of vortex-skyrmions.  The pinning of skyrmions would also restore  superfluidity.

There are many open questions related to  the LIM state. In particular, how rigid is the $\hat\textbf{l}$-texture at large length scales and how strongly are the vortex-skyrmions pinned by aerogel? 
The problem of rigidity  in the LIM state is thirty years old, see the paper by Efetov and Larkin \cite{EfetovLarkin} and recent publications  \cite{Itakura}. 
The idea of quasi-long-range order (QLRO) with  a power-law decay of correlators discussed for a vortex lattice in superconductors \cite{Giamarchi} is applicable to $^3$He-A in aerogel.  Probably  the QLRO does not occur  in isotropic aerogel because of the non-Abelian $SO(3)$ group of triad
$\hat\textbf{m}$, $\hat\textbf{n}$, $\hat\textbf{l}$, but it should take place in a stretched aerogel where  the easy plane anisotropy  reduces  $SO(3)$ to its Abelian $U(1)\times U(1)$ subspace. Following Ref. \cite{Nattermann} one may expect that the power law is determined by relations  between all 7  stiffness parameters $K$ in the $^3$He-A gradient energy \cite{VollhardtWolfle}.
 The phase transition from LIM state with QLRO to the disordered LIM state or to the non-superfluid LIM state  can be generated by increasing the density of skyrmions.
The states $c$ and $f$  in NMR experiments \cite{Dmitriev} and states in hydrodynamic experiments \cite{Fisher} (see below) may differ by density of skyrmions and thus may have different superfluid properties. 

\begin{figure}
\centerline{\includegraphics[width=1.0\linewidth]{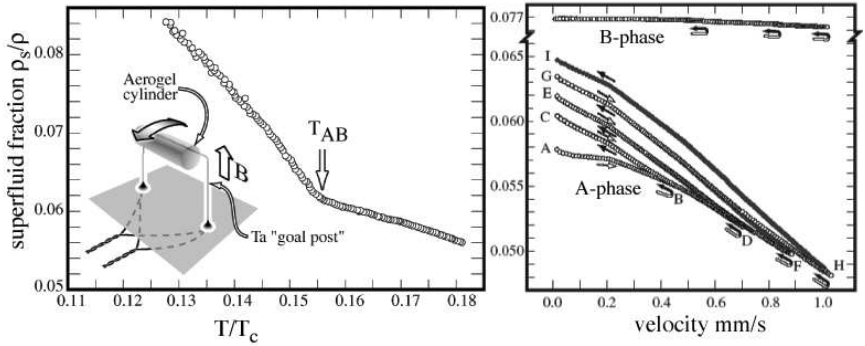}}
\caption{Anisotropic superfluid density of $^3$He-A in aerogel according to Ref.  \protect\cite{Fisher}. 
Superfluid density is history dependent and depends on the applied superfluid velocity. }  
\label{fig:Pickett} 
\end{figure}

The external  superflow  influences  the LIM state due to anisotropy of the superfluid density  in Eq.(\ref{Anisotropy}): since $\rho_{s\parallel}<\rho_{s\perp}$,  the applied $\textbf{v}_s$ produces the global easy-axis anisotropy for $\hat\textbf{l}$. This suggests that the sufficiently large $\textbf{v}_s$  destroys the LIM state, which is consistent with hydrodynamic experiment   \cite{Fisher} demonstrating   glassy behavior of  $\hat\textbf{l}$-texture at low $\textbf{v}_s$ and more regular behavior at large $\textbf{v}_s$. If the initial state $A$ in  Fig.  \ref{fig:Pickett} is fully disordered but rigid, its superfluid density must be $\rho_s^A=(1/3)\rho_{s\parallel}+(2/3)\rho_{s\perp}$.
  With increasing  $\textbf{v}_s$  the state becomes more and more uniform.  
In  the state $H$, which is reached at large  $\textbf{v}_s$, the vector $\hat\textbf{l}$ is oriented along $\textbf{v}_s$, and one has
 $\rho_s^H= \rho_{s\parallel}<\rho_s^A$. Because of the hysteresis, the uniform texture persists when velocity is reduced,  but  now with $\hat\textbf{l}\perp \textbf{v}_s$ as dictated by geometry of the sample; as a result in  
the subsequent state I  one has  $\rho_s^I= \rho_{s\perp}>\rho_s^A$.  

Another open problem is the role of extended objects. (i) Topological defects with hard cores  (single-quantum and half-quantum vortices) may be  pinned by aerogel. Trapping of singular vortices by aerogel has been observed in $^3$He-B \cite{KunimatsuQFS}. This suggests that  $c$ and $f$ states may have different abundance of defects.  (ii) The network of connected aerogel strands may have the long-range correlations (e.g. due to elastic deformations of aerogel). The interplay of  the quenched long-range and short-range  disorder may also lead to two LIM states with different scales \cite{Fedorenko}.

 The LIM effect is rather subtle in $^3$He-A, it is destroyed by weak anisotropy produced by deformation of aerogel and/or by flow. This suggests that maybe the adequate description of the LIM effect  is  in terms of  statistics of random deformations $u_{ik}$ of aerogel with the energy term in Eq.(\ref{RegularAnisotropyEnergy}), instead of the discussed statistics of random orientations of strands.  On the other hand, the global anisotropy produced by sufficiently large deformations allows us to study novel phenomena which cannot be observed in bulk  $^3$He-A. For example, orientation of  $\hat\textbf{l}$ along the magnetic field $\textbf{H}$  (not possible in a pure bulk $^3$He-A)  stabilizes: (i)  Alice strings (half-quantum vortices) in rotating sample \cite{SalomaaVolovik};  and (ii) the phase-coherent precession of magnetization  \cite{BunkovVolovik} recently observed in a squeezed aerogel \cite{JapCQS,KunimatsuQFS}.  The coherent  precession in $^3$He-A is another  realization of Bose-Einstein condensation (BEC) of magnons. The first example of magnon BEC found in 1984 in $^3$He-B is known as the homogeneously precessing domain (HPD, see reviews \cite{BECReview,QFSBEC}), and historically this was the first BEC which was experimentally stabilized. Magnon condensates in $^3$He-A and condensates in $^3$He-B (HPD, the state HPD$_2$ observed in a stretched aerogel \cite{HPD2} and $Q$-balls \cite{Q}) all have different spin-superfluid properties \cite{QFSBEC}.

In conclusion, $^3$He-A in aerogel is one of the most interesting objects for investigation of the Larkin-Imry-Ma and related effects.

\begin{acknowledgements}
I thank Yu.M. Bunkov, V.V. Dmitriev, T. Giamarchi, T. Nattermann and Y. Tanaka for illuminating discussions,   the Academy of Finland and  the Russian Foundation for
Fundamental Research (grant 06-02-16002-a) for support.
\end{acknowledgements}


\end{document}